\documentclass[preprint2]{emulateapj}
\usepackage{graphicx}
\usepackage{bm}
\usepackage{txfonts}
\usepackage{url}
\graphicspath{{./fig/}{./png/}}

\newcommand{\EQ}{\begin{equation}}
\newcommand{\EE}{\end{equation}}
\newcommand{\EQA}{\begin{eqnarray}}
\newcommand{\EEA}{\end{eqnarray}}

\newcommand{\pd}{\partial}

\newcommand{\DIV}{\bm{\nabla} \cdot }
\newcommand{\CURL}{\bm{\nabla} \times }

\newcommand{\mean}[1]{\overline{#1}}
\newcommand{\meanv}[1]{\overline{\bm #1}}

\newcommand{\etat}{\eta_{\rm t}}

\newcommand{\urms}{u_{\rm rms}}

\newcommand{\brms}{B_{\rm rms}}
\newcommand{\Beq}{B_{\rm eq}}

\newcommand{\kef}{k_{\rm f}}

\newcommand{\Pm}{{\rm Pm}}
\newcommand{\Rm}{{\rm Rm}}
\newcommand{\Rey}{{\rm Re}}
\newcommand{\Pra}{{\rm Pr}}
\newcommand{\Ta}{{\rm Ta}}
\newcommand{\Ra}{{\rm Ra}}

\newcommand{\Co}{{\rm Co}}

\def\onethird{{\textstyle{1\over3}}}
\def\onehalf{{\textstyle{1\over2}}}

\begin{document}

\title{Large-scale dynamos in rigidly rotating turbulent convection}

\author{Petri J.\ K\"apyl\"a and Maarit J.\ Korpi}
\affil{Observatory, University of Helsinki, PO Box 14,
FI-00014 University of Helsinki, Finland}
\email{petri.kapyla@helsinki.fi
($ $Revision: 1.138 $ $)
}

\author{Axel Brandenburg}
\affil{NORDITA, AlbaNova University Center, Roslagstullsbacken 23,
SE-10691 Stockholm, Sweden}

\begin{abstract} 
  The existence of large-scale dynamos in rigidly rotating
  turbulent convection without shear is studied using three-dimensional 
  numerical simulations of penetrative rotating compressible convection.
  We demonstrate that rotating convection in a Cartesian domain can drive a
  large-scale dynamo even in the absence of shear. The large-scale field 
  contains a significant fraction of the total field in the saturated 
  state. 
  The simulation results are compared with
  one-dimensional mean-field dynamo models where turbulent transport 
  coefficients, as determined using the test field method, are used.
  The reason for the absence of large-scale dynamo action in earlier studies
  is shown to be due to the rotation being too slow: whereas the
  $\alpha$-effect can change sign, its magnitude stays approximately
  constant as a function of rotation, and the turbulent diffusivity
  decreases monotonically with increasing rotation. Only when rotation 
  is rapid enough a large-scale dynamo can be excited. 
  The one-dimensional mean-field model with dynamo coefficients from
  the test field results predicts reasonably well the dynamo 
  excitation in the direct simulations. This result further 
  validates the test field procedure and reinforces the interpretation 
  that the observed dynamo is driven by a turbulent $\alpha$-effect.
  This result demonstrates the existence of an $\alpha$-effect and
  an $\alpha^2$-dynamo with natural forcing.
\end{abstract} 

\keywords{MHD -- convection -- turbulence -- Sun: magnetic fields -- stars: magnetic fields}


\section{Introduction}
Convective instability drives turbulence in the outer layers of
late-type stars, such as the Sun. The large-scale magnetic fields of
these stars are thought to arise from the interaction of turbulent
convection and the overall rotation of the object. In mean-field
dynamo theory (e.g.\ Moffatt 1978; Parker 1979; Krause \& R\"adler
1980; R\"udiger \& Hollerbach 2004), this process
relies on large-scale differential
rotation ($\Omega$-effect) producing toroidal field by shearing and
the $\alpha$-effect which regenerates the poloidal field. In simple
situations, the $\alpha$-effect is related to the kinetic helicity of
the flow (e.g.\ Steenbeck \& Krause 1969).
Dynamos where differential rotation is important, e.g.\ the
solar dynamo, are therefore often called $\alpha\Omega$-dynamos.

The $\alpha\Omega$ dynamo process was first invoked by Parker
(1955) to explain solar magnetism. Mean-field models have been
used extensively ever since to study various aspects of
dynamos. Although very useful in their own right, these models often
rely on ill-known parameterizations of turbulence, such as the
$\alpha$-effect and turbulent diffusivity. Only during recent years
numerical simulations have reached a level of sophistication where
large-scale $\alpha\Omega$-dynamos have been obtained
self-consistently within the framework of local Cartesian simulations.
These simulations operate in highly simplified
situations and the turbulence is driven by external forcing (e.g.\
Brandenburg et al.\ 2001; Brandenburg \& K\"apyl\"a
2007; K\"apyl\"a \& Brandenburg 2009). In rapidly
rotating stars differential rotation is likely to play only a minor
role (e.g.\ Hall 1991) and there the $\alpha$-effect generates also the toroidal
field. These systems are called $\alpha^2$-dynamos. Again,
such solutions have been found from direct simulations of forced
turbulence (Brandenburg 2001; Mitra et al.\ 2009b,c).

On the other hand, numerical simulations of magnetoconvection have
been around for at least two decades, but somewhat surprisingly
large-scale dynamos were not found until quite recently (Jones \&
Roberts 2000; Rotvig \& Jones 2002; Browning et al.\ 2006; Brown et
al.\ 2007; K\"apyl\"a et al.\ 2008, hereafter Paper I; Hughes \&
Proctor 2009).
The first two studies are related to the geodynamo and therefore
convection is rotationally dominated. The next two
use a global models with an imposed tachocline (Browning et al.\ 2006) or
rapid rotation (Brown et al.\ 2007), respectively. The
last two are local models with imposed shear flows, reminiscent of
the shear dynamos reported from non-helical turbulence (Yousef et al.\
2008a,b; Brandenburg et al.\ 2008). The origin
of large-scale fields in the non-helically forced simulations
cannot be due to the mean-field $\alpha$-effect. However, the incoherent
$\alpha$-shear dynamo (e.g.\ Vishniac \& Brandenburg 1997; Proctor 2007) 
or the mean-field shear-current effect (Rogachevskii \&
Kleeorin 2003, 2004; Kleeorin \& Rogachevskii 2008) could drive these 
dynamos. In the shearing
local convection simulations the same two effects have been suggested
mainly due to the fact that rotating convection alone has so far
been found unable to generate large-scale magnetic fields 
(e.g.\ Nordlund et al.\ 1992; Brandenburg et al.\ 1996; 
Cattaneo \& Hughes 2006; Tobias et al.\ 2008),
although it should produce a mean $\alpha$-effect according to theory.
This drawback has prompted speculations that the $\alpha$-effect
in its mean-field incarnation simply does not work
(Hughes \& Proctor 2009).

A conclusive proof of the existence of an $\alpha^2$-dynamo should be the
demonstration of large-scale dynamo action in a simulation without shear.
An additional property of such a setup is that the Vishniac \& Cho (2001)
flux of magnetic helicity is absent.
This might lead to slow saturation of the large-scale magnetic field,
unless there are other ways of shedding small-scale magnetic helicity.

The most natural way of explaining the large-scale magnetic
fields seen in the rapidly rotating global convection simulations
(Brown et al.\ 2007) would be in terms of a mean-field
$\alpha^2$- or $\alpha\Omega$-dynamo.
In the global models large-scale shear flows can also be generated, so it
is not straightforward to determine what type of dynamo is operating there.
In the present study, we use local simulations to
demonstrate that large-scale dynamos can indeed be excited in rapidly rotating
convection, i.e.\ in the absence of shear, as long as rotation is rapid enough.
We determine the turbulent transport coefficients from the simulations using
the test field procedure (Schrinner et al.\ 2005, 2007) and use them in a one-dimensional dynamo
model to test their consistency with the direct simulations. A more
thorough study of the turbulent transport coefficients from convection using the test field method
is presented elsewhere (K\"apyl\"a et al.\ 2009, hereafter Paper II).

\section{The model}
\label{sec:model}
Our model setup is similar to that used by Brandenburg et al.\
(1996), Ossendrijver et al.\ (2001, 2002), 
K\"apyl\"a et al.\ (2004, 2006), and those in Papers I and II.
A rectangular portion of a star is modeled
by a box situated at colatitude $\theta$. 
The box is divided into
three layers, an upper cooling layer, a convectively unstable layer,
and a stable overshoot layer (see below). The following set of
equations for compressible magnetohydrodynamics is being solved:
\begin{equation}
\frac{\partial \bm A}{\partial t} = {\bm U} \times {\bm B} -\eta\mu_0{\bm J}, \label{equ:AA}
 \end{equation}
\begin{equation}
\frac{\mathrm{D} \ln \rho}{\mathrm{D}t} = -\DIV{\bm U},
 \end{equation}
\begin{equation}
 \frac{\mathrm{D} \bm U}{\mathrm{D}t} = -\frac{1}{\rho}{\bm \nabla}p + {\bm g} - 2\bm{\Omega} \times \bm{U} + \frac{1}{\rho} \bm{J} \times {\bm B} + \frac{1}{\rho} \bm{\nabla} \cdot 2 \nu \rho \mbox{\boldmath ${\sf S}$}, \label{equ:UU}
 \end{equation}
\begin{equation}
 \frac{\mathrm{D} e}{\mathrm{D}t} = - \frac{p}{\rho}\DIV {\bm U} + \frac{1}{\rho} \bm{\nabla} \cdot K \bm{\nabla}T + 2 \nu \mbox{\boldmath ${\sf S}$}^2 + \frac{\eta}{\rho} \mu_0\bm{J}^2 - \frac{e\!-\!e_0}{\tau(z)}, \label{equ:ene}
 \end{equation}
where $\mathrm{D}/\mathrm{D}t = \pd/\pd t + \bm{U} \cdot \bm{\nabla}$.
The magnetic field is written in terms of the magnetic vector potential,
$\bm{A}$, with $\bm{B} = \bm{\nabla} \times \bm{A}$,
$\bm{J} =\mu_0^{-1}\bm{\nabla} \times \bm{B}$ is the current density,
$\mu_0$ is the vacuum permeability, $\eta$ and $\nu$ are the magnetic 
diffusivity and kinematic viscosity, respectively, $K$ is the heat 
conductivity, $\rho$ is the density, $\bm{U}$ is the
velocity, $\bm{g} = -g\hat{\bm{z}}$ is the gravitational acceleration,
and $\bm{\Omega}=\Omega_0(-\sin \theta,0,\cos \theta)$ is the rotation vector.
The fluid obeys an ideal gas law $p=(\gamma-1)\rho e$, where $p$
and $e$ are pressure and internal energy, respectively, and
$\gamma = c_{\rm P}/c_{\rm V} = 5/3$ is the ratio of specific heats at
constant pressure and volume, respectively.
The specific internal energy per unit mass is related to the
temperature via $e=c_{\rm V} T$.
The rate of strain tensor $\mbox{\boldmath ${\sf S}$}$ is given by
\begin{equation}
{\sf S}_{ij} = \onehalf (U_{i,j}+U_{j,i}) - \onethird \delta_{ij} \DIV \bm{U}.
\end{equation}
The last term of Eq.~(\ref{equ:ene}) describes cooling at the top of
the domain. 
Here $\tau(z)$ is a cooling time which has a profile
smoothly connecting the upper cooling layer and the convectively
unstable layer below, where $\tau\to\infty$.

The positions of the bottom of the box, bottom and top of the
convectively unstable layer, and the top of the box, respectively,
are given by $(z_1, z_2, z_3, z_4) = (-0.85, 0, 1, 1.15)d$, where $d$ is 
the depth of the convectively unstable layer. Initially
the stratification is piecewise polytropic with polytropic indices
$(m_1, m_2, m_3) = (3, 1, 1)$, which leads to a convectively unstable
layer above a stable layer at the bottom of the domain. The cooling 
term leads to a stably stratified isothermal layer at the top.
The horizontal extent of the box $L_{\rm H}\equiv L_x=L_y$ is varied 
between $2d$ and $8d$. 
All simulations with rotation use $\theta=0\degr$ corresponding to
the north pole.

\subsection{Units and nondimensional parameters}
Nondimensional quantities are obtained by setting
\begin{eqnarray}
d = g = \rho_0 = c_{\rm P} = \mu_0 = 1\;,
\end{eqnarray}
where $\rho_0$ is the initial density at $z_2$. The units of length, time,
velocity, density, entropy, and magnetic field are
\begin{eqnarray}
&& [x] = d\;,\;\; [t] = \sqrt{d/g}\;,\;\; [U]=\sqrt{dg}\;,\;\; [\rho]=\rho_0\;,\;\; \nonumber \\ && [s]=c_{\rm P}\;,\;\; [B]=\sqrt{dg\rho
_0\mu_0}\;. 
\end{eqnarray}
We define the fluid and magnetic Prandtl numbers and the Rayleigh
number as
\begin{eqnarray}
\Pra=\frac{\nu}{\chi_0}\;,\;\; \Pm=\frac{\nu}{\eta}\;,\;\; \Ra=\frac{gd^4}{\nu \chi_0} \bigg(-\frac{1}{c_{\rm P}}\frac{{\rm d}s}{{\rm d}z
} \bigg)_0\;,
\end{eqnarray}
where $\chi_0 = K/(\rho_{\rm m} c_{\rm P})$ is the thermal
diffusivity, and $\rho_{\rm m}$ is the density in the middle of
the unstable layer, $z_{\rm m} = \onehalf(z_3-z_2)$. The entropy 
gradient, measured at $z_{\rm m}$, in the nonconvecting hydrostatic state,
is given by
\begin{eqnarray}
\bigg(-\frac{1}{c_{\rm P}}\frac{{\rm d}s}{{\rm d}z}\bigg)_0 = \frac{\nabla-\nabla_{\rm ad}}{H_{\rm P}}\;,
\end{eqnarray}
where $\nabla-\nabla_{\rm ad}$
is the superadiabatic temperature gradient with 
$\nabla_{\rm  ad} = 1-1/\gamma$, $\nabla = (\pd \ln T/\pd \ln
  p)_{z_{\rm m}}$, and where $H_{\rm P}$ is the pressure scale height
(Brandenburg et al.\ 2005).
The amount of stratification is determined by the parameter 
$\xi_0 =(\gamma-1) e_0/(gd)$, which is the pressure scale height at
the top of the domain normalized by the depth of the unstable layer.
We use $\xi_0 =1/3$ in all cases,
which results in a density contrast of about 23 across the domain.
We define the fluid and magnetic Reynolds numbers via
\begin{eqnarray}
{\rm Re} = \frac{\urms}{\nu \kef}\;,\;\; {\rm Rm} = \frac{\urms}{\eta \kef} = \Pm\ {\rm Re}\;,
\end{eqnarray}
where $\kef = 2\pi/d$ is adopted as an estimate
for the wavenumber of the energy-carrying eddies.
Note that with our definition $\Rm$ is smaller than the
usual one by a factor $2\pi$.
The amount of rotation is quantified by
\begin{eqnarray}
{\rm Co} = \frac{2\Omega_0}{\urms \kef}\;. \label{equ:Co}
\end{eqnarray}
In order to facilitate comparison with earlier studies of rotating
Rayleigh-B\'enard convection we also quote the value of the Taylor number,
\begin{equation}
\Ta=\left(2\Omega_0 d^2/\nu\right)^2=\left(4\pi^2\Co\,\Rm/\Pm\right)^2.
\end{equation}
The equipartition magnetic field is defined by 
\begin{equation}
\Beq \equiv \langle\mu_0\rho\bm{U}^2\rangle^{1/2},\label{equ:Beq}
\end{equation}
where angular brackets denote volume averaging.

\subsection{Boundary conditions}

\begin{deluxetable*}{cccccccccccccc}
\tabletypesize{\scriptsize}
\tablecaption{Summary of the runs. 
  Here, $\mbox{\rm Ma}=\urms/(gd)^{1/2}$, $\tilde{B}_{\rm rms} \equiv
  \brms/B_{\rm eq}$, and the $\tilde{B}^{(k)}$ are the sum of the rms-values
  of the Fourier amplitudes of the horizontal components of the magnetic
  field for mode $k$ normalized by $\brms$.
  Values with $\tilde{B}^{(0)}\ge0.07$ or $\tilde{B}^{(1)}\ge0.15$
  are shown in bold face and indicate the presence of significant
  large-scale fields.
  The last column denotes the magnetic field boundary
  condition at the $z$-boundaries.}
\tablewidth{0pt}
\tablehead{
\colhead{Run} & \colhead{grid} & \colhead{$L_{\rm H}$} & \colhead{$\Pra$} & 
\colhead{$\Pm$} & \colhead{$\Ra$} & \colhead{$\Rm$} & \colhead{$\Co$} & 
\colhead{$\mbox{Ta}$} & \colhead{$\mbox{Ma}$} &
\colhead{$\tilde{B}_{\rm rms}$} & \colhead{$\tilde{B}^{(0)}$} & 
\colhead{$\tilde{B}^{(1)}$} & \colhead{BC}
}
\startdata
A1   & $128^3$  & $2$ & $0.27$ &  $2$  & $1.5\cdot10^6$ & $70$  & $0$    & $0$            & $0.044$ & $0.12$ & $0.02$ & $0.05$ & vf \\ 
A2   & $128^3$  & $2$ & $0.27$ &  $2$  & $1.5\cdot10^6$ & $68$  & $0.37$ & $2.5\cdot10^5$ & $0.043$ & $0.23$ & $0.04$ & $0.10$ & vf \\ 
A3   & $128^3$  & $2$ & $0.27$ &  $2$  & $1.5\cdot10^6$ & $68$  & $0.74$ & $10^6$         & $0.043$ & $0.25$ & $0.04$ & $0.09$ & vf \\ 
A4   & $128^3$  & $2$ & $0.27$ &  $2$  & $1.5\cdot10^6$ & $69$  & $1.5$  & $4\cdot10^6$   & $0.043$ & $0.14$ & $0.02$ & $0.05$ & vf \\ 
A5   & $128^3$  & $2$ & $0.27$ &  $2$  & $1.5\cdot10^6$ & $62$  & $4.1$  & $2.5\cdot10^7$ & $0.039$ & $0.18$ & $0.04$ & $0.07$ & vf \\ 
A6   & $128^3$  & $2$ & $0.27$ &  $2$  & $1.5\cdot10^6$ & $42$  & $11.6$ & $10^8$         & $0.028$ & $0.38$ &$\bm{0.22}$&$\bm{0.18}$& vf \\ 
A7   & $128^3$  & $2$ & $0.27$ &  $1$  & $1.5\cdot10^6$ & $22$  & $11.6$ & $10^8$         & $0.028$ & $0.27$ &$\bm{0.15}$&$\bm{0.16}$& vf \\ 
A8   & $128^3$  & $2$ & $0.27$ & $0.67$& $1.5\cdot10^6$ & $17$  & $9.8$  & $10^8$         & $0.033$ & $0.00$ & $0.00$ & $0.00$ & vf \\ 
A9   & $128^3$  & $2$ & $0.27$ & $0.4$ & $1.5\cdot10^6$ & $10$  & $9.7$  & $10^8$         & $0.033$ & $0.00$ & $0.00$ & $0.00$ & vf \\ 
A10  & $256^3$  & $2$ & $0.14$ & $2$ & $3.0\cdot10^6$ & $107$  & $9.5$  & $4\cdot10^8$        & $0.034$ & $0.41$ & $\bm{0.08}$ & $\bm{0.15}$ & $vf$ \\ 
           \hline
B1   & $256^2\times128$ & $4$ & $0.27$ & $2$ & $1.5\cdot10^6$ & $77$  & $0$    & $0$            & $0.048$ & $0.22$ & $0.02$ & $0.06$ & vf \\ 
B2   & $256^2\times128$ & $4$ & $0.27$ & $2$ & $1.5\cdot10^6$ & $67$  & $0.38$ & $2.5\cdot10^5$ & $0.042$ & $0.25$ & $0.02$ & $0.05$ & vf \\ 
B3   & $256^2\times128$ & $4$ & $0.27$ & $2$ & $1.5\cdot10^6$ & $66$  & $0.76$ & $10^6$         & $0.042$ & $0.22$ & $0.02$ & $0.05$ & vf \\ 
B4   & $256^2\times128$ & $4$ & $0.27$ & $2$ & $1.5\cdot10^6$ & $67$  & $1.5$  & $4\cdot10^6$   & $0.042$ & $0.16$ & $0.01$ & $0.03$ & vf \\ 
B5   & $256^2\times128$ & $4$ & $0.27$ & $2$ & $1.5\cdot10^6$ & $55$  & $4.6$  & $2.5\cdot10^7$ & $0.035$ & $0.31$ & $0.05$ & $0.09$ & vf \\ 
B6   & $256^2\times128$ & $4$ & $0.27$ & $2$ & $1.5\cdot10^6$ & $42$  & $11.9$ & $10^8$       & $0.027$ & $0.41$ &$\bm{0.12}$&$\bm{0.15}$& vf \\ 
B7   & $256^2\times128$ & $4$ & $0.27$ & $2$ & $1.5\cdot10^6$ & $41$  & $12.4$ & $10^8$       & $0.026$ & $0.48$ &$\bm{0.18}$&$\bm{0.26}$& pc \\ 
           \hline
C1   & $512^2\times128$ & $8$ & $0.27$ & $2$ & $1.5\cdot10^6$ & $42$  & $12.0$ & $10^8$       & $0.026$ & $0.45$ &$\bm{0.07}$&$\bm{0.17}$& vf \\ 
           \hline
D1   & $512^2\times256$ & $4$ & $0.14$ & $2$ & $3.0\cdot10^6$ & $97$  & $10.5$ & $4\cdot10^8$ & $0.030$ & $0.58$ &$\bm{0.08}$&$\bm{0.30}$& vf 
\enddata
\label{tab:runs}
\end{deluxetable*}

Stress-free boundary conditions are used in the vertical ($z$) 
direction for the velocity,
\begin{equation}
U_{x,z} = U_{y,z} = U_z = 0,
\end{equation}
where commas denote partial derivatives,
and either vertical field of perfect conductor conditions are used 
for the magnetic field, i.e.\
\begin{eqnarray}
B_x = B_y &=& 0 \;\; \mbox{(vertical field)},\\
B_{x,z} = B_{y,z} = B_z &=& 0 \;\; \mbox{(perfect conductor)}.
\end{eqnarray}
The vertical field (VF) conditions permit a magnetic helicity flux, but no
Poynting flux, whereas the perfect conductor (PC) conditions do not allow
helicity fluxes either.
In the $x$ and $y$ directions
periodic boundary conditions are used. The simulations were made with
the {\sc Pencil Code}%
\footnote{\texttt{http://www.nordita.org/software/pencil-code/}},
which uses sixth order explicit finite differences in space and third
order accurate time stepping method. Resolutions of up to $512^2\times256$
mesh points were used.

\begin{figure*}[t]
\centering
\includegraphics[width=\textwidth]{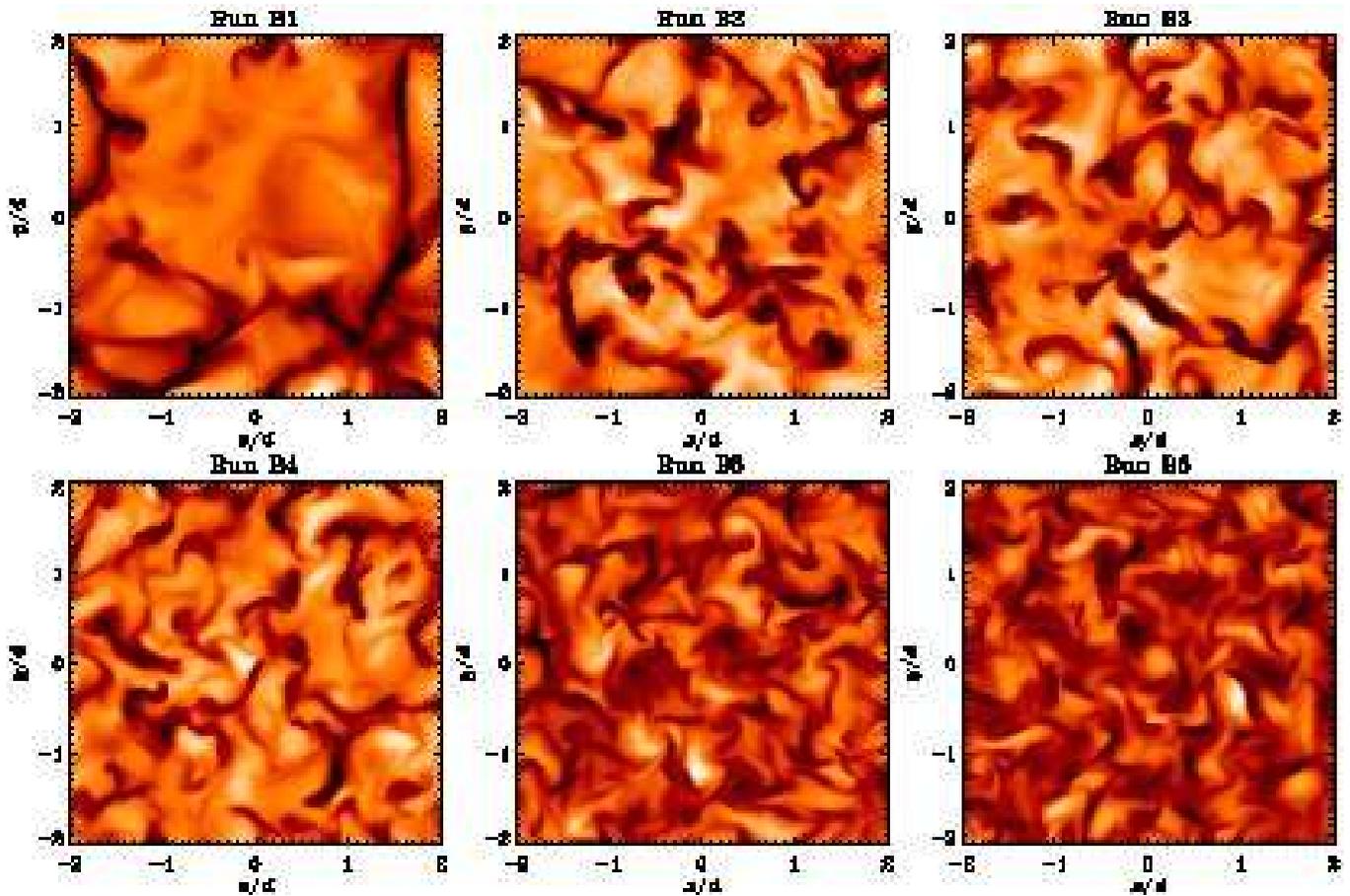}
\caption{Vertical velocity $U_z$  from the middle of the convectively
  unstable layer from Runs~B1-B6 in the kinematic stage at 
  $t=400\sqrt{d/g}$, corresponding to $t \urms \kef \approx 70\ldots110$,
  depending on the run.
  Light (dark) color indicates ascending (descending) motion.}
\label{fig:puz_rot}
\end{figure*}

\section{Results}
\label{sec:results}
We perform a parameter study where we vary the system size,
the effect of rotation, quantified by $\Co$, and the Reynolds numbers
in order to study the existence of large-scale dynamos in rotating
convection. The
runs are listed in Table~\ref{tab:runs}. We briefly describe the
hydrodynamics of these runs in \S~\ref{sec:hydro}, and the
results on dynamo excitation and large-scale magnetic fields
are given in
\S~\ref{sec:magnetic}. Finally, in \S~\ref{sec:transport} and
\S~\ref{sec:mf} the turbulent transport coefficients and the
interpretation of the results in the framework of mean-field
electrodynamics, respectively, are presented.

\subsection{The hydrodynamic state}
\label{sec:hydro}
All our simulations start with a small (of the order of $10^{-5}\Beq$) 
random magnetic field.
This field remains dynamically insignificant in the initial stages,
which, in most cases, span the first 200 turnover times of the run.
As in Paper I we consider this interval to represent the
hydrodynamic state of the simulation.

The effect of rotation on convection has been described in various
papers (e.g.\ K\"apyl\"a et al.\ 2004; Giesecke et al.\ 2005,
Cattaneo \& Hughes 2006; Hughes \& Cattaneo 2008; Tobias
et al.\ 2008). The most visible effect is the decreasing size of
convective cells as rotation is increased; see
Fig.~\ref{fig:puz_rot}. Without rotation (Run~B1) and for horizontal
extent $L_{\rm H}/d=4$ the flow is dominated by essentially a single
large cell, but already for $\Co\approx1.5$ (Run~B4) there are of the
order of ten or more cells in the domain. When rotation is increased
further, the cell size continues to diminish.
This feature is familiar from simulations of unstratified convection
(King et al.\ 2009), although there the flow appears to become more
easily laminar at large Taylor numbers ($\Ta=10^8$) and modestly
large Rayleigh numbers ($\Ra=1.5\cdot10^6$) that we used here
(see Table~\ref{tab:runs}).

\begin{figure}[t]
\centering
\includegraphics[width=\columnwidth]{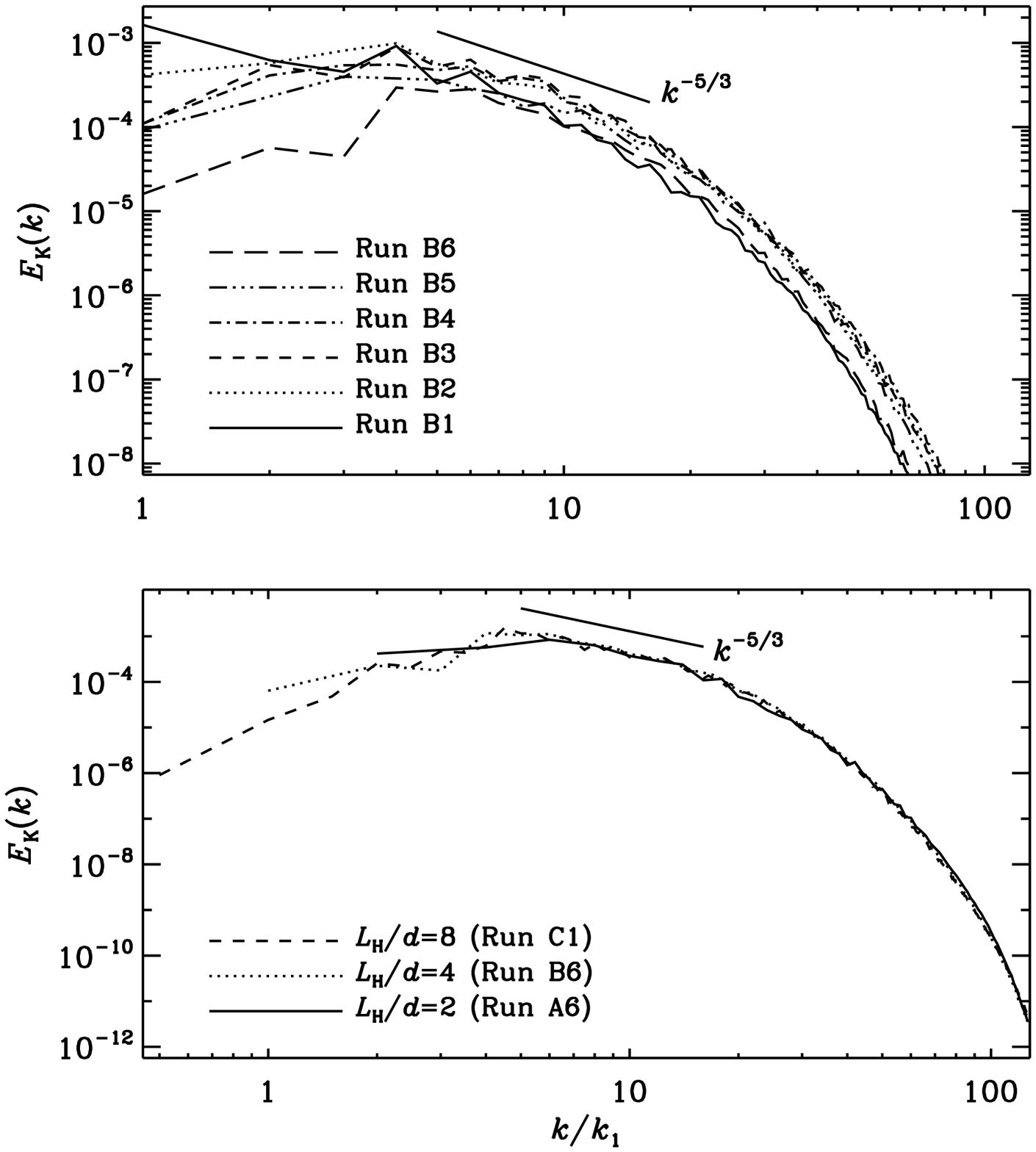}
\caption{Two-dimensional power spectra of velocity from the middle of
  the convectively unstable layer as functions of rotation (upper panel)
  and system size (lower panel) in the kinematic regime at 
  $t=400\sqrt{d/g}$. Power law slopes
  proportional to $k^{-5/3}$ are shown for reference. The spectra in
  the lower panel have been shifted so that the same spatial scales
  coincide on the $x$-axis and scaled by $L_{\rm H}/d$ so
  that the curves lie on top of each other.}
\label{fig:pspecu_kin}
\end{figure}

The decrease in cell size
is also ma\-ni\-fes\-ted by two-dimensional power spectra of the
velocity, see the upper panel of Fig.~\ref{fig:pspecu_kin}. For
the nonrotating case (Run~B1) the most power is found at $k/k_1=1$,
where $k_1=2\pi/L_{\rm H}$. There is a tendency
for the wavenumber of the maximum, $k_{\rm max}$, to shift towards 
higher $k$ for more rapid rotation.
The lower panel of Fig.~\ref{fig:pspecu_kin} shows the velocity power
spectra for Runs~A6, B6, and C1 with $L_{\rm H}/d=2,4,8$,
respectively, shifted so that the same spatial scales
coincide. All runs have $\rm{Re}\approx28$ (21) and $\Co\approx9$ (12) 
in the kinematic (saturated) regime. For
the smallest system size the energy still peaks near the box
scale. When the box size increases the spatial scale at which most
energy is found stays the same and, as a consequence, the scale
separation, $k_{\rm max}/k_1$, between the energy carrying scale and
the box scale increases.

Our finding that $k_{\rm max}$ is independent of
the horizontal system size is important because large-scale
dynamos require some amount of scale separation, i.e.\ that the power
spectrum of the turbulence should peak at some scale smaller than the
system size. This allows the energy to cascade to smaller and larger
scales. The former eventually dissipates the energy into heat whereas
the latter can generate large-scale structures through an inverse
cascade of energy. The turbulent $\alpha$-effect can also be
interpreted as an inverse cascade (Brandenburg 2001). If, however, the
flow energy peaks at, or close to, the scale of the system, it is impossible
to cascade energy to larger scales.

\begin{figure}[t]
\centering
\includegraphics[width=\columnwidth]{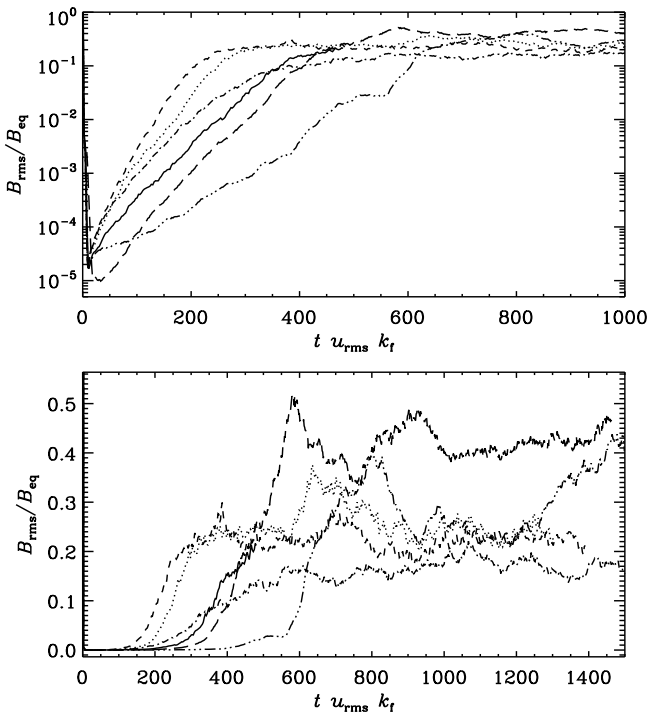}
\caption{Upper panel: root mean square of the total magnetic field as a function of time for 
the same runs as in the upper panel of Fig.~\ref{fig:pspecu_kin}.
Lower panel: the same as above but in linear scale.}
\label{fig:pbrmsco}
\end{figure}

\begin{figure}[t]
\centering
\includegraphics[width=\columnwidth]{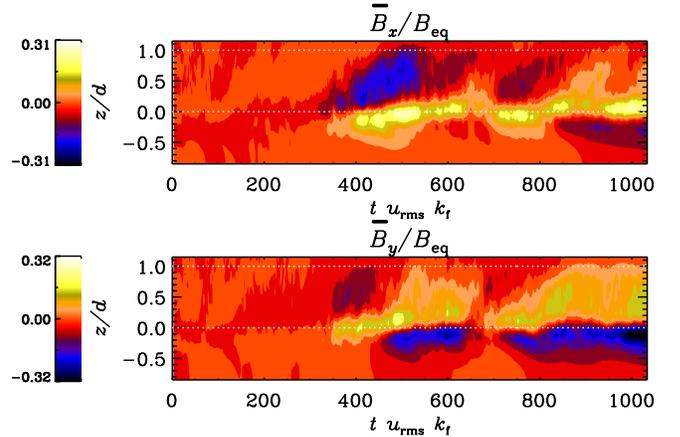}
\caption{Horizontally averaged magnetic field components $\mean{B}_x$
  (top panel) and $\mean{B}_y$ (bottom panel), normalized by the
  equipartition magnetic field strength. From Run~A6 with
  $\Co\approx12$ and $\Rm\approx42$.
  The horizontal dotted lines at $z/d=0$ and $z/d=1$ indicate the base and
  top of the convectively unstable layer, respectively.}
\label{fig:st_128b}
\end{figure}

\begin{figure*}[t]
\centering
\includegraphics[width=\textwidth]{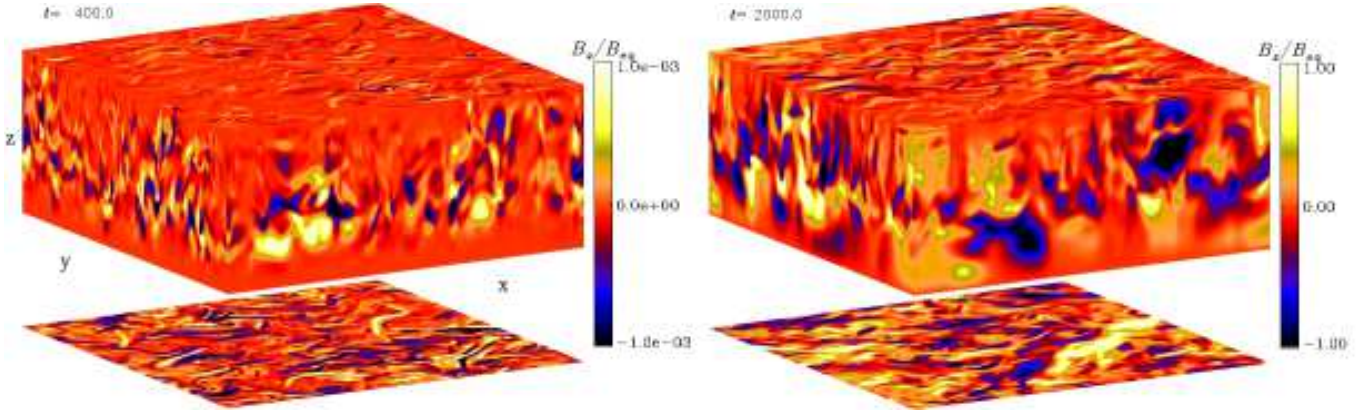}
\caption{Magnetic field component $B_x$ for Run~D1 in the kinematic
  (left panel) and saturated (right) states. The sides of the box show
  the periphery of the domain whereas the top and bottom panels show
  the field from the top ($z=d$) and bottom ($z=0$) of the
  convectively unstable layer, respectively. See also 
  \texttt{http://www.helsinki.fi/\ensuremath{\sim}kapyla/movies.html}.}
\label{fig:boxplot_256x128a}
\end{figure*}

\begin{figure}
\centering
\includegraphics[width=\columnwidth]{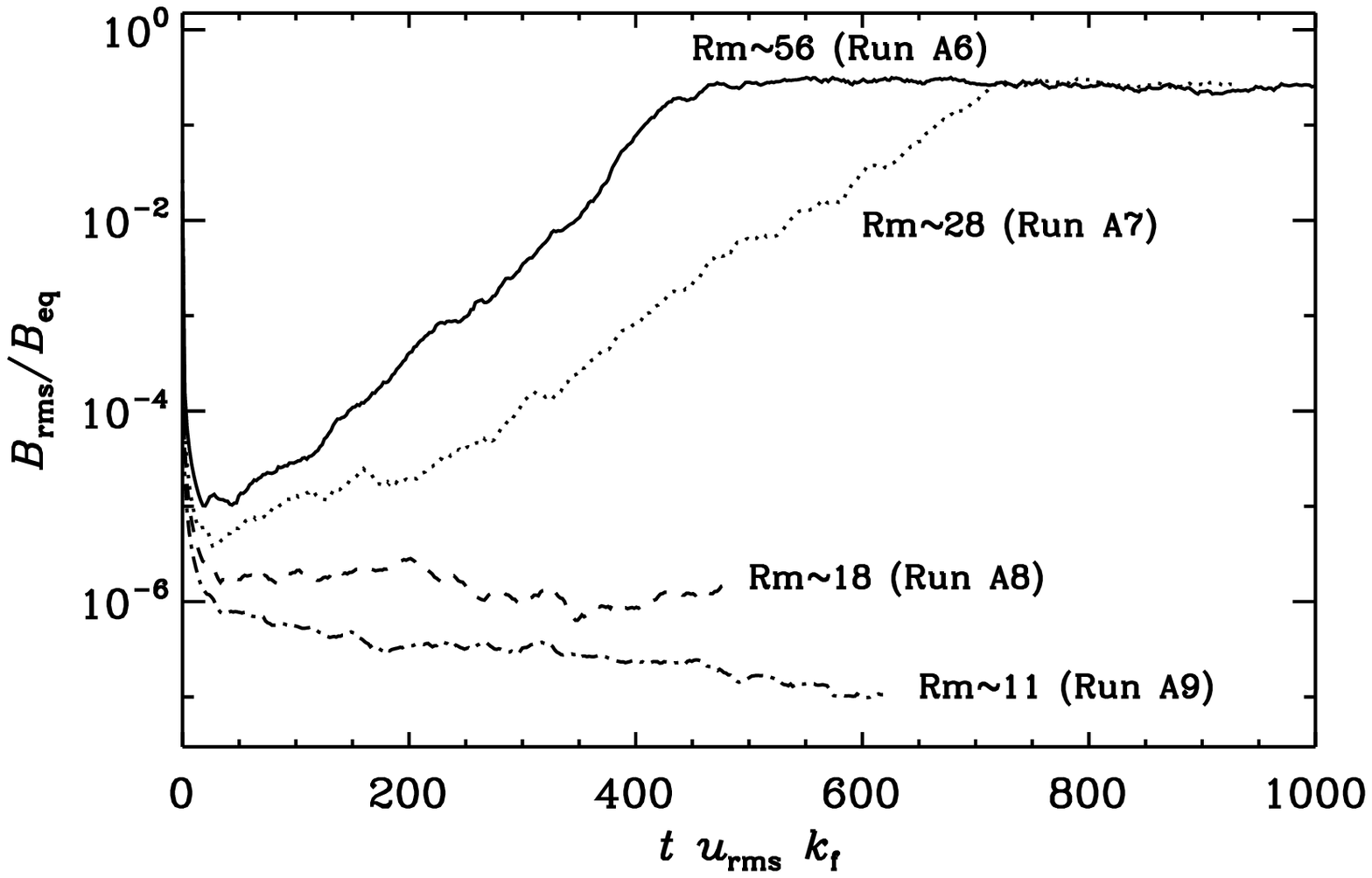}
\caption{Root mean square total magnetic field as a function of time for 
four magnetic Reynolds numbers for Runs~A6--A9.}
\label{fig:pbrmsrm}
\end{figure}

\subsection{Dynamo excitation and large-scale magnetic fields}
\label{sec:magnetic}
Recent numerical studies have indicated that shear plays an important
role in exciting large-scale dynamos (e.g.\ Yousef et al.\
2008a,b; Paper I; Hughes \& Proctor 2009).
Although the process generating the large-scale magnetic field in the
nonhelical case is still under debate (e.g.\ Rogachevskii \& Kleeorin
2003, 2004; R\"udiger \& Kitchatinov 2006; Proctor 2007; Brandenburg
et al.\ 2008; Schekochihin et al.\ 2009; Sridhar \& Subramanian 2009), 
there are some indications
that in helical turbulence with shear a classical $\alpha\Omega$, or
$\alpha$-shear dynamo, simultaneously with a shear-current and 
$\bm{\Omega}\times{\bm J}$-dynamos, might be the explanation 
(Paper II). However, in
the absence of shear, large-scale dynamos due to helical turbulence
have so far been obtained only in idealized systems where the helical
flow is driven by external forcing (Brandenburg 2001; Mitra et al.\ 2009b,c). In particular,
the lack of large-scale dynamos in rotating convection has been
puzzling (e.g.\ Cattaneo \& Hughes 2006),
although an $\alpha$-effect should be present according to theory. 
One possible explanation is that, whereas the turbulence in more
idealized studies can be almost fully helical owing to the forcing, in
rotating convection the fractional helicity is often much smaller
(e.g.\ Brandenburg et al.\ 1990, 1996; K\"apyl\"a et al.\ 2004).

In Paper II we found that the turbulent $\alpha$-effect increases and
turbulent diffusivity decreases, respectively, as functions of
rotation for small fluid Reynolds numbers (see below). 
These results suggest that a large-scale dynamo in rotating convection 
is excited
only if rotation is rapid enough. Our conjecture is that this regime
might not have been reached in earlier studies thus failing to produce
large-scale dynamos.

The evolution of the rms-value of the total magnetic field for
Runs~B1--B6 is shown in Fig.~\ref{fig:pbrmsco}. We find that the
growth rate of the field increases for $\Co\la0.74$ (Runs~B1--B3)
after which it starts to decrease. However, for the most rapid
rotation (Run~B6) the growth rate again increases and is close to the
nonrotating case. The saturation level of the magnetic field is
practically constant for $\Co\la0.74$ (Runs~B1--B3) and somewhat lower
for $\Co\approx1.5$ (Run~B4). For the two most rapidly rotating cases
the saturation level is generally higher and more variable. In these runs the
maxima of the field are associated with periods where significant
amounts of large-scale fields are present; see Fig.~\ref{fig:st_128b}
for the time evolution of horizontally averaged fields $\mean{B}_x$ and
$\mean{B}_y$ for Run~A6. Comparing the field in the kinematic and
saturated stages (Fig.~\ref{fig:boxplot_256x128a}) shows that in the
kinematic regime the field is concentrated in small scales, whereas in
the late stages a clear large-scale structure is visible.
The behavior of the growth rate and
saturation level of the field as functions of rotation could be
understood as follows: the small-scale dynamo is enhanced for slow
rotation but starts to be rotationally quenched for $\Co\ga1$.
This might be explicable by an associated decrease in the length scale
of the turbulence.
On the other hand, for
rapid enough rotation, i.e.\ $\Co\ga3$, the large-scale dynamo becomes
excited and increases the growth rate and saturation level.
We stress that the proposed suppression of the small-scale dynamo action
due to rotation is at this stage only a conjecture that is
consistent with the numerical data. Whether this interpretation is
correct should be studied in more detail separately.
It should also be noted that we cannot expect rotation to quench the
small-scale dynamo in the astrophysically relevant regime where $\Rm$
is much larger than in the present simulations. There are, however, no
compelling reasons to expect that the large-scale dynamo is suppressed
when a small-scale dynamo is also present as long as magnetic helicity
fluxes are allowed to leave the domain.
For the most rapidly rotating case, with $\Co\approx9$ in the
kinematic regime, the critical magnetic Reynolds number is roughly 20,
see Fig.~\ref{fig:pbrmsrm}. In Paper I we reported that in the absence
of rotation, the critical $\Rm$ for the excitation of a small-scale
dynamo is roughly 30 for a similar system.

\begin{figure}[t]
\centering
\includegraphics[width=\columnwidth]{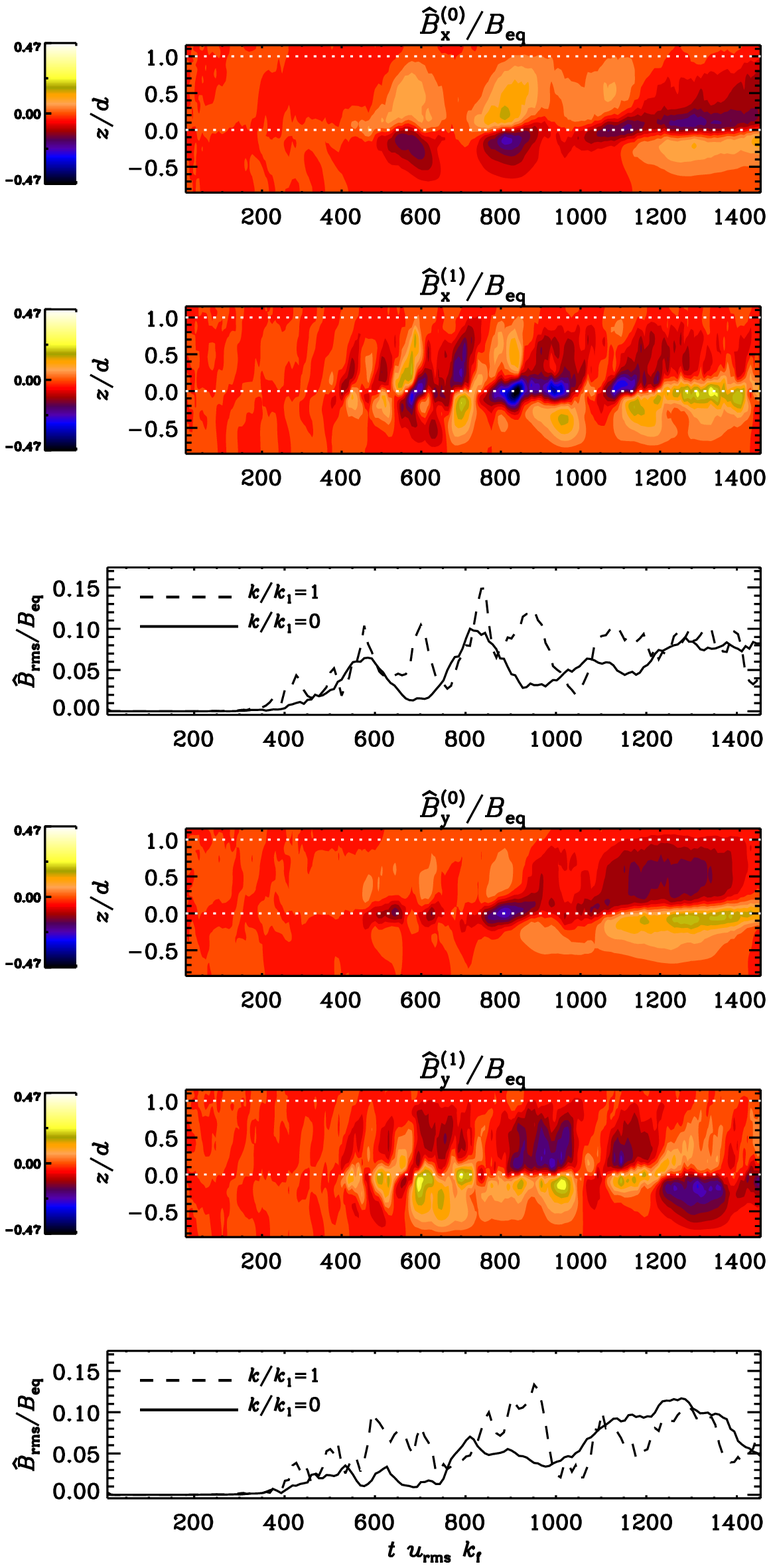}
\caption{Fourier amplitudes of the two horizontal components of the
  magnetic field for the two smallest wavenumbers $k/k_1=0$ and 1 from 
  Run~B6. From
  top to bottom: $\hat{B}_x^{(0)}(z,t)$, $\hat{B}_x^{(1)}(z,t)$, their
  rms-values; $\hat{B}_y^{(0)}(z,t)$, $\hat{B}_y^{(1)}(z,t)$, and
  their rms-values.}
\label{fig:pfamp_256a}
\end{figure}

In Paper I we found the large-scale field to be dominated by the 
$k_x=k_y=0$ mode and 
it could thus be well described by
a horizontal average.
However, in the present case the mean fields show a more 
complicated structure and tend to have comparable contributions
from $k_x$, $k_y\neq 0$ modes as well (see, e.g.\ the right panel 
of Fig.~\ref{fig:boxplot_256x128a}).
Thus a simple horizontal average no longer
suffices to represent the large-scale field. For a better measure, we
perform a two-dimensional Fourier transform of the field at each
horizontal layer, the result of which we denote by $\hat{B}_i(k_x,k_y,z,t)$. 
Now we project 
this quantity onto a one-dimensional wavenumber $k^2=k_x^2+k_y^2$
and denote its Fourier amplitude by $\hat{B}_i^{(m)}(z,t)$, where
$m$ is the discretized wavenumber bin corresponding to $k/k_1$. So, for
$k/k_1=0$, $\hat{B}_i^{(0)}(z,t)$ is the same as a horizontal average
$\mean{B}_i$. For larger $k$, $\hat{B}_i^{(m)}(z,t)$ gives a measure
of the strength of the mode $k$. 
In the following we restrict the
analysis to the smallest values of $k/k_1=0\ldots2$ that describe the 
large-scale field.

\begin{figure}[t]
\centering
\includegraphics[width=\columnwidth]{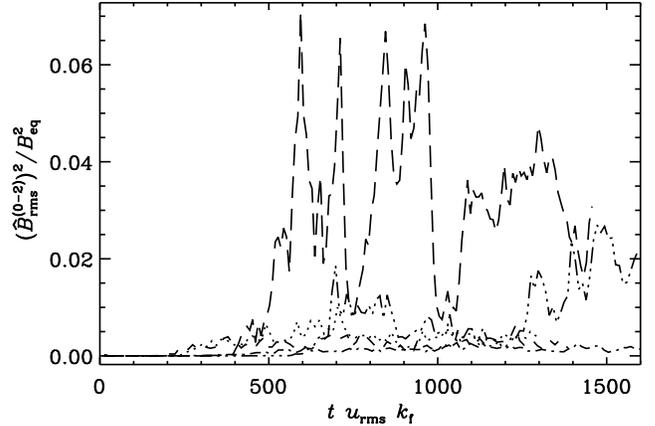}
\caption{The sums of the root mean square values of the Fourier
  amplitudes of $B_x$ and $B_y$ for the modes $k/k_1=0\ldots2$ as
  function of rotation for the Runs~B1-B6. Linestyles as in the upper
  panel of Fig.~\ref{fig:pspecu_kin}}
\label{fig:pmean}
\end{figure}

\begin{figure}[t]
\centering
\includegraphics[width=\columnwidth]{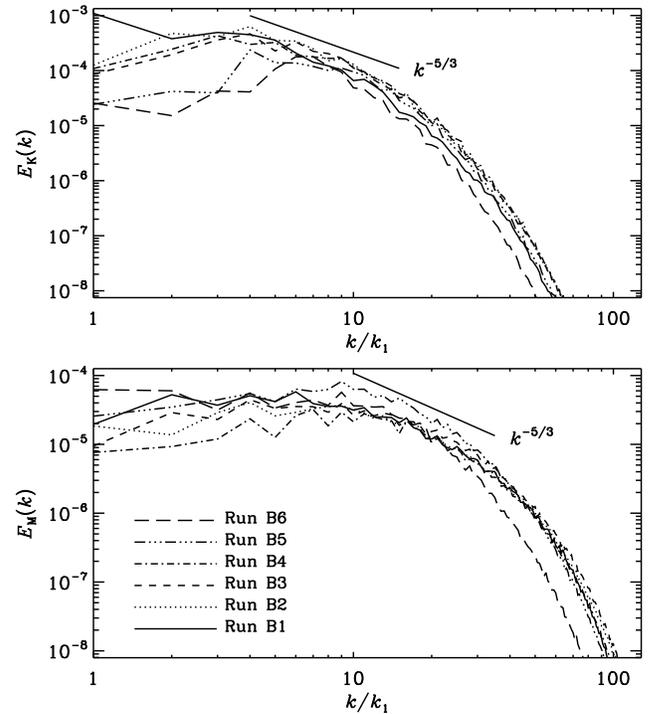}
\caption{Two-dimensional power spectra of velocity (upper panel) and
  magnetic field (lower panel) from Runs~B1-B6 in the saturated
  state. Power laws proportional to $k^{-5/3}$ are shown for
  reference.}
\label{fig:pspec_rot}
\end{figure}

Representative results for the horizontal components of
$\hat{B}_i^{(0)}(z,t)$ and $\hat{B}_i^{(1)}(z,t)$ 
from Run~B6 with $\Co\approx12$ are shown in
Fig.~\ref{fig:pfamp_256a}. The large-scale field seems to show
opposite signs in the convectively unstable and stable layers.
Sign changes also occur but they are rather irregular and do not appear
to follow a consistent cycle. The root mean square of the
amplitudes for the $k/k_1=0$ and $1$ contributions to the horizontal
magnetic fields are of the order of ten per cent of the equipartition
value. This is to be contrasted with Fig.~\ref{fig:pmean} where the sums
of the rms-values of the amplitudes of the three smallest wavenumbers 
are shown as functions of
rotation. Substantial large-scale fields are observed only for the two
largest values of $\Co$; for $\Co\la1.5$ the runs show very weak
large-scale contributions. This is also visible in the two-dimensional
power spectra, taken from the middle of the convectively unstable
layer; see Fig.~\ref{fig:pspec_rot}. The
most rapidly rotating case (Run~B6) is the only one showing clear signs of
large-scale fields, in accordance with the fact that the large-scale field
is only periodically present in Run~B5. The velocity spectra in the
saturated state are similar to those in the kinematic phase.

In comparison to earlier studies of rotating convection (e.g.\
Cattaneo \& Hughes 2006; Tobias et al.\ 2008), we note that it is
characteristic of these studies that when the Rayleigh number is
increased, the Taylor number is kept constant. Increasing $\Ra$ in
these models generates a larger $\urms$ and this inevitably means that
the Coriolis number decreases as the Rayleigh number is increased,
i.e.\ for a fixed $\Ta$ the rotational influence is large for a small
Rayleigh number and vice versa. For example, in the paper of Cattaneo
\& Hughes (2006), the smallest Rayleigh numbers in combination with
$\Ta=5\cdot10^5$ gives a Coriolis number comparable to our largest
values. However, these simulations do not exhibit dynamo action due to
a too low $\Rm$, whose value also depends on $\urms$. For their
highest Rayleigh number case, however, $\Rm$ is large enough for
dynamo excitation but the Coriolis number is smaller by approximately
an order of magnitude and no large-scale fields are observed.
Similar arguments apply to the simulations of Tobias et al.\ (2008).

\begin{figure}[t]
\centering
\includegraphics[width=\columnwidth]{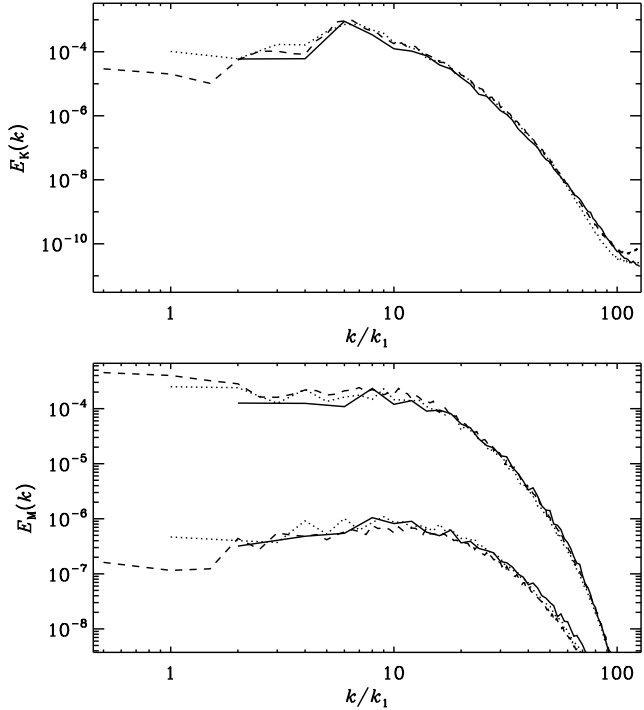}
\caption{Two-dimensional power spectra of velocity (upper panel) and
  magnetic field (lower panel) as functions of system size. Linestyles
  and scaling as in the lower panel of Fig.~\ref{fig:pspecu_kin}. In the lower panel
  the upper curves show the spectra from the saturated state whereas the
  lower curves show the spectra from the kinematic state multiplied by $10^7$.}
\label{fig:pspec_size}
\end{figure}

\begin{figure}[t]
\centering
\includegraphics[width=\columnwidth]{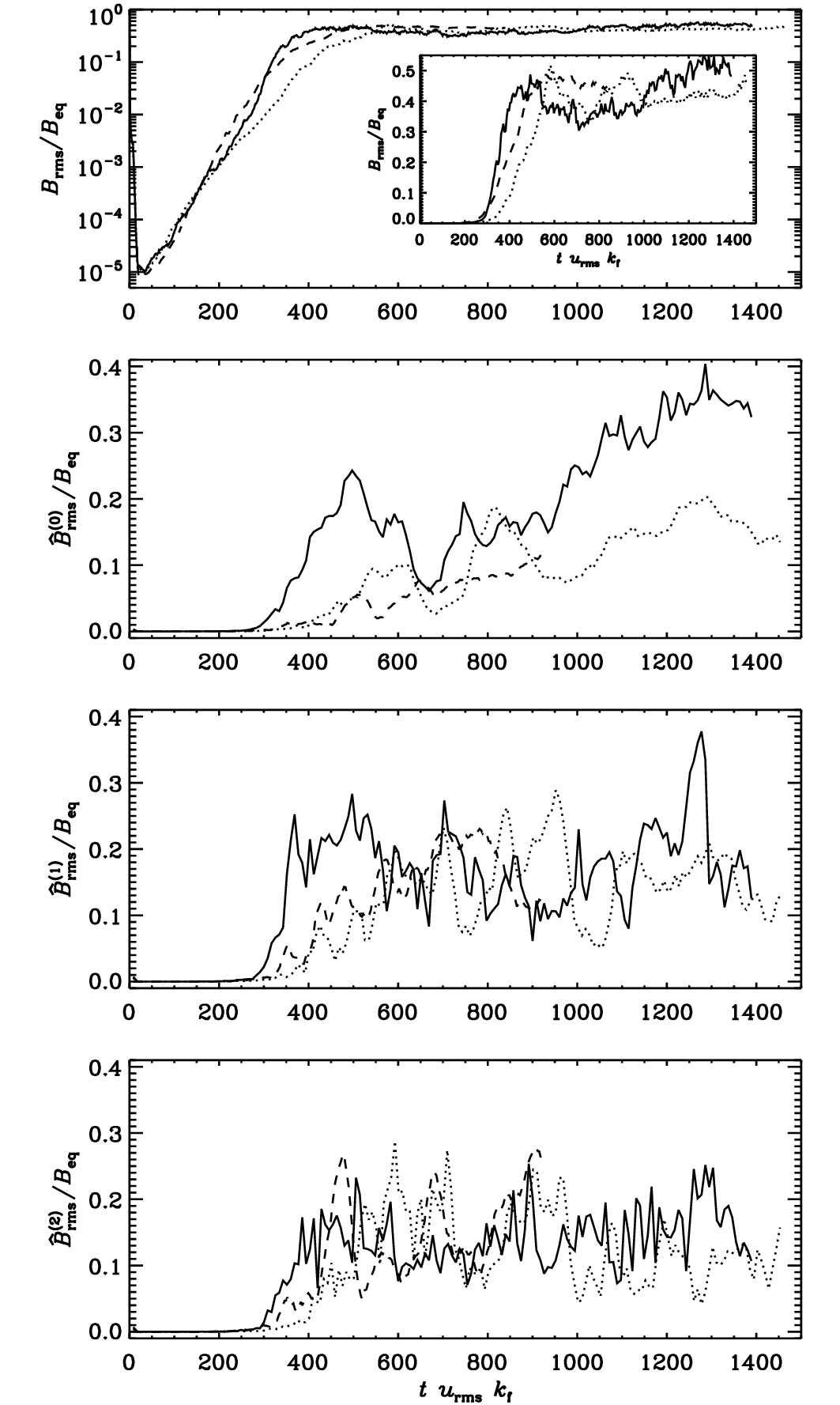}
\caption{Top panel: rms-values of the total magnetic field as
  functions of the system size. The three lower panels show the sums
  of the rms-values of the Fourier amplitudes of $B_x$ and $B_y$ for
  $k/k_1=0$ (second panel from the top), $k/k_1=1$ (second from the
  below), and $k/k_1=2$ (bottom panel). Linestyles as in
  the lower panel of Fig.~\ref{fig:pspecu_kin}.}
\label{fig:pbrmssize}
\end{figure}

We find that the large-scale dynamo is excited for all box sizes for
the most rapidly rotating case explored in the present study, as is
evident from the spectra shown in Fig.~\ref{fig:pspec_size}. From the
spectra it would seem that an increasing amount of energy is in the
large scales as the box size increases. The growth rate of the total
field does not show any clear trend with the system size: the largest
departure from a constant growth rate is the somewhat lower value for
Run~B6 with the intermediate box size (see Fig.~\ref{fig:pbrmssize}).

Figure \ref{fig:pbrmssize} shows the sums of the Fourier amplitudes of
the three smallest wavenumbers as functions of the system size. For
the smallest box (Run~A6), the large-scale field is more concentrated
on the $k/k_1=0$ contribution, whereas in Run~B6 with $L_{\rm H}/d=4$
the amplitudes for $k/k_1=0$ and 1 are similar. For the largest domain
size the $k=0$ mode is significantly weaker than the $k/k_1=1$ and 2
modes. These results suggest that for the present parameters the
maximum size of the large-scale structures is somewhere in the range
$2<L_{\rm max}/d<8$.

Comparing the saturation level of the large-scale magnetic field in
the small box Runs~A7, A6, and A10 shows a significant decrease in the
$m=0$ component in Run~A10 whereas the strength of the $m=1$ mode is
only mildly affected. On the other hand, comparing Runs~B6 and D1 with
a larger domain size shows again a decreasing $m=0$ contribution in
Run~D1 but a two times larger $m=1$ component. However, these numbers
should be taken only as a rough guide because the large-scale
contribution to the magnetic field shows large fluctuations and the
higher $\Rm$ runs are fairly short. Taken at face value, the results
would seem to suggest that the strength of the $m=0$ mode decreases
with increasing $\Rm$ and that the $m=1$ mode remains unaffected or
that it can even increase. We note that the highest $\Rm$ runs also
have larger fluid Reynolds and Rayleigh numbers which means that also
the flow is more turbulent in those cases which could affect the
dynamo and thus the saturation level of the large-scale field.

\begin{figure}[t]
\centering
\includegraphics[width=\columnwidth]{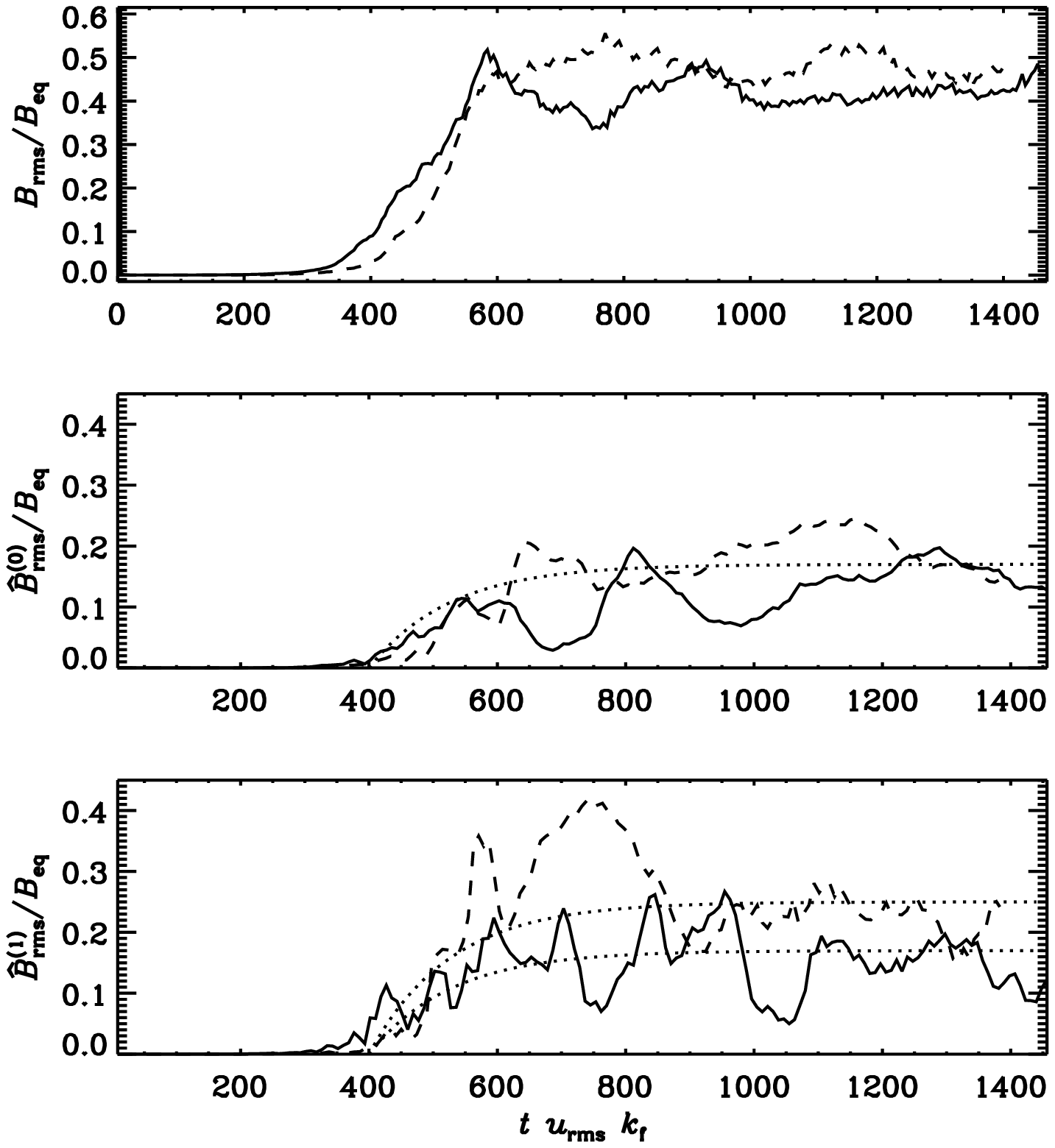}
\caption{Top panel: rms-values of the total magnetic field as
  functions of time for vertical field (Run~B6, solid lines)  
  and perfect conductor boundary conditions (Run~B7, dashed
  lines). The two lower panels show the sums of the rms-values of
  the Fourier amplitudes of $B_x$ and $B_y$ for $k/k_1=0$ (middle
  panel) and $k/k_1=1$ (bottom panel). The dotted lines in the two
  lower panels show a saturation predictor according to the model of
  Brandenburg (2001).}
\label{fig:psatb}
\end{figure}

Although we have used open (vertical field) boundary conditions
that do permit magnetic helicity fluxes, such fluxes may not actually
occur unless they are driven toward the boundaries by internal magnetic
helicity fluxes.
One such flux is the Vishniac-Cho (2001) flux, but it requires shear
which is absent in our case.
Other fluxes are possible (Subramanian \& Brandenburg 2006), but we do
not know how efficient they are in our model.
It is therefore unclear whether one should expect the saturation of
the large-scale field to occur on a dynamical or a resistive time scale,
and what the relevant length scale is.
In Fig.~\ref{fig:psatb} we show the saturation behavior for Run~B6
and compare with a curve proportional to $1-\exp[-2\eta k^2(t-t_{\rm s})]$
that would be expected for resistively dominated saturation behavior
(Brandenburg 2001).
Here, $k=\kef$ has been chosen and
$t_{\rm s}$ marks the end of the linear growth phase, which is
also the time when the small-scale magnetic field has saturated.
The result is not entirely conclusive, and larger magnetic Reynolds
number would be needed to clarify this, but it is certainly possible that
saturation of the large-scale field is resistively dominated.
Similar behavior is certainly expected in the case of perfect conductor
boundary conditions (Run~B7; see the dashed lines in Fig.~\ref{fig:psatb}). 
We note that the saturation level of the total and mean fields are
higher in Run~B7 with perfect conductor boundary conditions than in
Run~B6 with vertical field conditions. The behavior is qualitatively
similar to the forced turbulence simulations of Brandenburg (2001).

\subsection{Turbulent transport coefficients}
\label{sec:transport}
We attempt to connect the dynamos seen in the direct simulations to
theoretical considerations by comparing the results to a mean-field
dynamo that relies on turbulent transport coefficients, such as the
$\alpha$-effect and turbulent diffusivity. 
In order to extract these
coefficients we use the test field method (Schrinner et al.\ 2005,
2007) in the kinematic regime. Detailed descriptions of the method,
as implemented here,
have been presented elsewhere (e.g.\ Brandenburg et al.\ 2008; Sur et
al.\ 2008; Mitra et al.\ 2009a; Paper II); here we only outline the
general procedure. The method relies on a set of orthogonal test
fields that do not evolve in time and do not react back onto the
flow. These properties allow an unambiguous determination of the
turbulent transport coefficients in the kinematic regime without the 
complications associated with some other methods that can be used.

We restrict the study to test fields that depend only on $z$
\begin{eqnarray}
\meanv{B}^{1c}&=&B_0(\cos kz,0,0), \quad \meanv{B}^{2c}=B_0(0,\cos kz,0),\\
\meanv{B}^{1s}&=&B_0(\sin kz,0,0), \quad \meanv{B}^{2s}=B_0(0,\sin kz,0),
\end{eqnarray}
where $B_0$ is the amplitude of the field and $k$ is the wavenumber of
the test field.
(The results are strictly independent of the value of $B_0$,
so in non-dimensional units it can be set to one.)
We use $k/k_1=1$, where $k_1=2\pi/L_z$, in all models.
The $k$-dependence of the coefficients in convection was studied in
Paper II. The electromotive force can be written as
\begin{equation}
\mean{\mathcal{E}}_i=\alpha_{ij}\mean{B}_j -\eta_{ij}\mu_0\mean{J}_j,
\end{equation}
where $\eta_{i1}= \eta_{i23}$ and $\eta_{i2}= -\eta_{i13}$. The 4+4
coefficients are then obtained by inverting a simple matrix equation,
relating the rank-2 tensor components to rank-3 tensor components.

It is convenient to discuss the results in terms of the quantities
\begin{eqnarray}
\alpha&=&\onehalf(\alpha_{11}+\alpha_{22}), \ \ \gamma=\onehalf(\alpha_{21}-\alpha_{12}), \\ \etat&=&\onehalf(\eta_{11}+\eta_{22}), \ \ \delta= \onehalf(\eta_{21}-\eta_{12}).
\end{eqnarray}
These quantities can be understood to represents the generation of
large-scale magnetic fields ($\alpha$), turbulent pumping of magnetic
fields ($\gamma$), turbulent diffusivity ($\etat$),
and the $\bm{\Omega}\times\bm{J}$ effect ($\delta$).
To normalize our results, we use isotropic expressions derived under
the first order smoothing approximation (see, e.g.\ Paper II)
\begin{eqnarray}
\alpha_0=\onethird \urms, \ \ \eta_{\rm t0}=\onethird \urms \kef^{-1}.
\end{eqnarray}

\begin{figure}[t]
\centering
\includegraphics[width=\columnwidth]{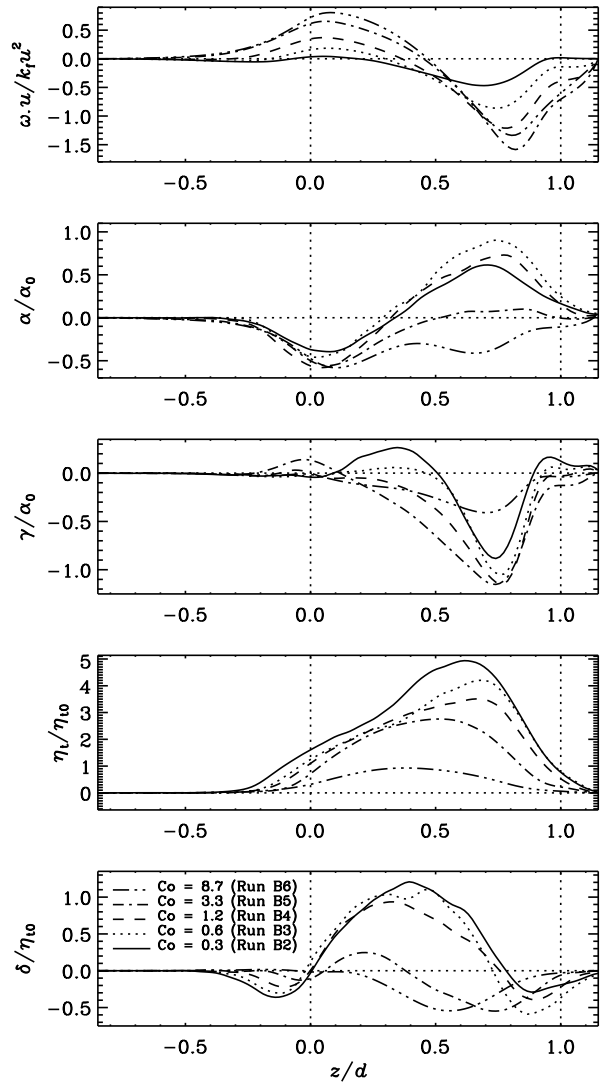}
\caption{From top to bottom: normalized profiles of
  kinetic helicity, $\alpha$, $\gamma$, $\etat$, and $\delta$ from
  kinematic test field simulations. The vertical dotted lines at 
  $z/d=0$ and $z/d=1$ indicate the base and top of the convectively
  unstable layer, respectively.}
\label{fig:palpeta}
\end{figure}

Figure \ref{fig:palpeta} shows the kinetic helicity 
$\overline{\bm{\omega}\cdot\bm{u}}$, where $\bm{\omega=\CURL\bm{u}}$, 
and the turbulent transport coefficients
for Runs~B2--B6 as functions of rotation. The magnetic Reynolds number 
varies between $\Rm\approx58\ldots86$ with larger values occurring for 
slower rotation. $\Pm=2$ used in all runs.
These simulations differ from those in Paper II in
that $\Pr$ and $\Pm$ are respectively five times and 2.5 times smaller. We
find that the kinetic helicity increases monotonically as a function
of rotation as in Paper II, although here the increase is not as steep
as in the previous results. The $\alpha$-effect is approximately
constant for $\Co\approx0.3\ldots1.2$, whereas for $\Co\approx3.3$
the positive region in the top half of the convectively unstable layer
disappears giving nearly zero value. For the most rapid rotation,
$\Co\approx8.7$, $\alpha$ is negative in the whole convection zone.
At first sight this result seems to contradict
the corresponding results of Paper II, but
we find that there is a qualitative change in the behavior of
$\alpha$ in the rapid rotation regime when the fluid Reynolds number
is increased. For $\Rey\approx2$ there is no sign change and a
monotonously increasing magnitude whereas for $\Rey\approx7$ the
positive region in the upper layers of the convectively unstable layer
disappears. Increasing the Reynolds number further to roughly 30 gives
the result shown in Fig.~\ref{fig:palpeta}.
We note that controlling $\Rey$ and $\Co$ a priori is difficult in the
rapid rotation regime because convection is increasingly suppressed 
especially for small Reynolds numbers.

For the slowest rotation, the turbulent pumping, $\gamma$, is very
similar in profile and magnitude as in Paper II. In the case of
$\gamma$ the behavior seems again different with negative values in
the whole convection zone for $\Co\ga1.17$. The magnitude of $\gamma$
is significantly decreased for our case with the most rapid rotation.
The turbulent diffusivity is rotationally quenched similarly as in Paper II.
Also the coefficient $\delta$ behaves in a similar fashion as in Paper II.

\begin{figure}[t]
\centering
\includegraphics[width=\columnwidth]{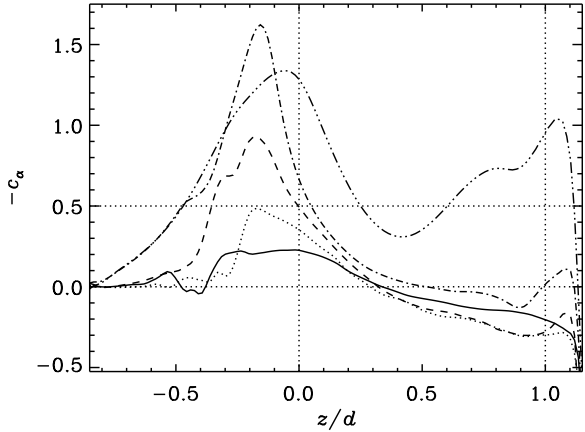}
\caption{Vertical profiles of $c_\alpha(z)$ for the same runs as in
  Fig.~\ref{fig:palpeta}. The horizontal line at $-c_\alpha=0.5$
  denotes where $|C_\alpha| = |c_\alpha|\kef L_z/2\pi=1$.}
\label{fig:pcalp}
\end{figure}

The relatively unchanged magnitude of $\alpha$ and the significantly
reduced value of $\etat$ suggests that excitation of a mean-field
dynamo could be possible in our cases with the most rapid rotation. One way to
quantify this is to compute a local dynamo number
\begin{equation}
c_\alpha(z)=\frac{\alpha}{\eta_{\rm T}\kef},
\end{equation}
where $\eta_{\rm T}=\etat+\eta$. The results are shown in
Fig.~\ref{fig:pcalp}.
We find that the dynamo number peaks in the overshoot layer just below
$z/d=0$ because the profile of the $\alpha$-effect extends somewhat
deeper than that of turbulent diffusivity. The magnitude of the
maximum of $c_\alpha$ in the overshoot increases monotonically until
$\Co\approx3.3$ but the value in the convectively unstable layer
remains small. For the most rapidly rotating case, $c_\alpha(z)$
increases significantly also in the convectively unstable region due
to the rotationally quenched value of $\etat$.
In the homogeneous case, where $c_\alpha$ is constant, dynamo action
is possible when its value based on the lowest wavenumber in the domain
exceeds unity, i.e.\ $C_\alpha\equiv c_\alpha\kef L_z/2\pi>1$.
In our case, owing to the presence of the overshoot layers, $\kef L_z/2\pi=2$,
so one might expect that dynamo action is possible when the average
value of $c_\alpha(z)$ exceeds 0.5.
For our case with the largest rotation rate this is clearly the case
for most of the domain.
We can therefore conclude that dynamo action should be possible
in that case.
However, a more detailed comparison would require using test
fields with wavenumbers other than $k=k_1$, as was done in Paper II for
a case with slower rotation than here.
We should therefore not over-interpret our present comparisons.

The significance of an overshoot layer in connection with large-scale
dynamo action is not yet entirely clear.
Tobias et al.\ (2008) argue that its presence is unimportant for promoting
the generation of large-scale fields, while Browning et al.\ (2006) find
it to be crucial.
However, as discussed in the present paper, the rotation rate used in
Tobias et al.\ (2008) may be too slow for a successful large-scale dynamo.
Another issue might be the absence of shear-induced helicity fluxes 
(Vishniac \& Cho 2001), as discussed in Paper~I.
In Browning et al.\ (2006), on the other hand, the presence of shear
in the overshoot layer plays an important role, as is demonstrated by
the striking difference to earlier results of Brun et al.\ (2004)
without overshoot.
Another aspect of the problem is that in local convection simulations
the amount of overshoot decreases as rotation increases
(Ziegler \& R\"udiger 2003; K\"apyl\"a et al.\ 2004).

\subsection{Mean-field models}
\label{sec:mf}
In order to check how well the turbulent transport coefficients presented in the
previous section describe the excitation of large-scale dynamos in the
direct simulations, we use a one-dimensional mean-field dynamo model in
which the test-field results for $\alpha_{ij}$ and $\eta_{ij}$ are
used as input parameters. The model solves the equation
\begin{eqnarray}
\dot{\mean{A}}_i = \alpha_{ij} \mean{B}_j -(\eta_{ij}+\eta\delta_{ij}) \mu_0 \mean{J}_j,
\end{eqnarray}
where the dot refers to a time derivative. The mean magnetic field
and current density are given by
$\meanv{B}=(-\mean{A}_y',\mean{A}_x',0)$ and
$\mu_0\mean{J}=-\mean{A}_i''$, respectively.
Here primes denote $z$-derivatives.

The test field results for $\alpha_{ij}$ and $\eta_{ij}$ can now be
directly used in the mean-field model, leaving little freedom. 
We must, however, bear in mind the limitations of this simple
model. Firstly, the transport coefficients were determined for a
single value of $k$, whereas in the direct simulations a larger set of
wavenumbers are available. Secondly, the model is restricted to fields
that depend only on $z$, whereas the direct simulations indicate that
the large-scale field can also vary in the horizontal directions.

\begin{figure}[t]
\centering
\includegraphics[width=\columnwidth]{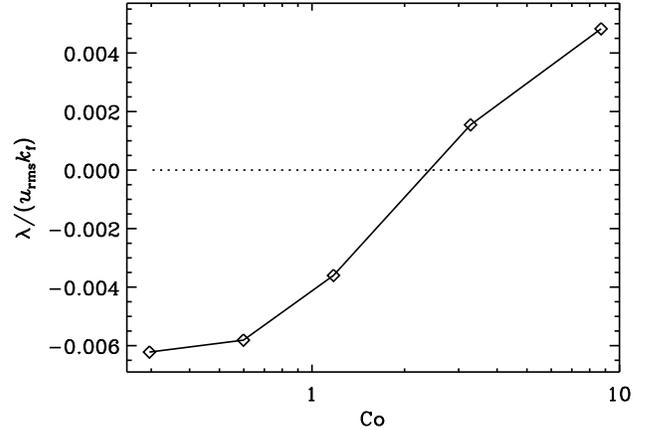}
\caption{Growth rate of the large-scale magnetic field for the same
  runs as in Fig.~\ref{fig:palpeta} from the one-dimensional dynamo
  model.}
\label{fig:pgr}
\end{figure}

Nevertheless, it is interesting to study whether the transport
coefficients derived using the test field procedure can excite a
dynamo in the present case. The results for the growth rate $\lambda$
of the magnetic field from the mean-field models, using the
coefficients shown in Fig.~\ref{fig:palpeta}, are shown in
Fig.~\ref{fig:pgr}. We find that the two most rapidly rotating runs 
show clear dynamo action
which is consistent with the direct simulations. These results support
the notion that the mechanism generating the large-scale magnetic
fields in the direct simulations is the turbulent $\alpha$-effect.

It should be noted that for values of $\lambda$ different from zero
the actual growth rate will not be exactly equal to $\lambda$, because
in general the components of $\alpha_{ij}$ and $\eta_{ij}$ depend on
frequency (and thereby on $\lambda$) -- just as they also depend on
the value of $k$; see Hubbard \& Brandenburg (2009) for details.


\section{Conclusions}
\label{sec:conclusions}
We use numerical simulations to demonstrate that rigidly rotating convection
can generate a large-scale dynamo. The importance of this result lies
in the fact that, according to mean-field theory, rotating
turbulent convection leads to an $\alpha$-effect, which should result
in a large-scale dynamo.
So far this has not been seen
in simulations without shear, leading to doubts of the
applicability of the mean-field approach.
The present results show that the lack of
large-scale dynamos in rotating convection previously reported is probably due to too
slow rotation. We find that an appreciable large-scale field is
generated in the direct simulations if $\Co\ga4$. 

With the smallest system size explored here, the scale separation is
small even in the most rapidly rotating case (Run~A6), i.e.\ $k_{\rm
  max}/k_1\approx2$. In this case the large-scale magnetic field is
concentrated in the $k=0$ mode. Doubling the domain size increases the
scale separation to $k_{\rm max}/k_1\approx5$ (Run~B6). Here the $k=0$
contribution is smaller and most of the magnetic energy is found
for $k/k_1=1$ and 2 modes. Doubling the size of the box once more, $k_{\rm
  max}/k_1\approx10$. The magnetic energy peaks at the largest scales,
although the $k=0$ mode is significantly weaker than in the cases with
smaller system sizes. This suggests that for the present parameters there is a maximum size for
the large-scale magnetic field structures which is larger than $2d$
and smaller than $4d$. 

We compute the turbulent transport coefficients using the test field
method and find that the magnitude of the $\alpha$-effect remains
relatively unaffected when rotation increases. However, at the same
time the turbulent diffusion is severely quenched by the rotation. When these
coefficients are used in a one-dimensional mean-field dynamo model,
the two most rapidly rotating runs exhibit a growing dynamo in
accordance with the direct simulations.
Although the mean-field model is too simple to fully describe the
field in the direct simulations, our results seem to validate the test
field method further and lend support to our interpretation that the
large-scale magnetic fields observed in the simulations are due to a
turbulent $\alpha$-effect.

On a more general note, the present results also represent a nice
demonstration of the usefulness of mean-field theory as a predictive tool.
In fact, as explained in \S~\ref{sec:magnetic}, our work was motivated by
the earlier findings of Paper~II that the $\alpha$-effect increases and
the turbulent diffusivity decreases as the rotation rate is increased.
This did already suggest the existence of an $\alpha^2$-dynamo for
sufficiently rapid rotation -- a suggestion that we have now been able
to confirm in this paper.

It should be noted that the magnetic field structure of mean-field
dynamos depends crucially on the geometry of the domain and the nature
of the boundary conditions.
Therefore, our results are not directly relevant to astrophysical bodies,
because their geometry is not Cartesian nor the boundaries periodic.
Our models can also not be thought of as local re\-pre\-sen\-ta\-tions of
a star, although it is possible to get some idea about the dependence
of the transport coefficients on latitude (Ossendrijver et al.\ 2002; 
K\"apyl\"a et al.\ 2006; Paper II).
Furthermore, the values of the magnetic Reynolds number are obviously
not in the astrophysically relevant regime.
Nevertheless, our simulations provide the first concrete evidence of dynamo
action from rotating convection without the additional help of shear.
This is an important point, because it proves for the first time that
an $\alpha^2$ dynamo from rotating convection exists and that it is
strong enough to produce large-scale fields.
This was thought impossible until now (Cattaneo \& Hughes 2006; 
Hughes \& Cattaneo 2008; Hughes \& Proctor 2009).

\acknowledgements
  The anonymous referee is acknowledged for the detailed comments. 
  The computations were performed on the facilities hosted by CSC - IT
  Center for Science Ltd. in Espoo, Finland, who are administered by
  the Finnish Ministry of Education.
  This research has benefitted from the computational resources granted 
  by the CSC to the grand challenge project ``Dynamo08''. 
  Financial support from the Academy of Finland grants No.\ 121431 (PJK)
  and 112020 (MJK) and the Swedish Research Council grant 621-2007-4064 (AB) is acknowledged.
  The authors
  acknowledge the hospitality of Nordita during the program
  `Turbulence and Dynamos' during which this work was initiated.

\end{document}